\documentclass[nofootinbib]{revtex4}

\usepackage[english]{babel}
\usepackage{graphicx}
\usepackage{psfrag}
\usepackage{amsmath}
\usepackage{amssymb}
\usepackage{verbatim}

\newcommand{\be}{\begin{equation}}
\newcommand{\ee}{\end{equation}}
\newcommand{\bea}{\begin{eqnarray}}
\newcommand{\eea}{\end{eqnarray}}
\newcommand{\bean}{\begin{eqnarray*}}
\newcommand{\eean}{\end{eqnarray*}}

\renewcommand{\b}{\langle}
\newcommand{\ket}{\rangle}

\newcommand{\irm}{{\rm i}}
\newcommand{\e}{{\rm e}}

\renewcommand{\d}{{\rm d}}
\newcommand{\cl}[1]{{\mathcal #1}}

\newcommand{\pa}{\partial}

\newcommand{\ts}{\textstyle}
\newcommand{\sst}{\scriptstyle}

\newcommand{\pdiff}[2]{\frac{\partial #1}{\partial #2}}

\newcommand{\bZ}{\mathbb{Z}}

\newcommand{\bR}{\mathbb{R}}

\newcommand{\clS}{\cl{S}}

\newcommand{\clJ}{\cl{J}}

\newcommand{\eq}[1]{(\ref{#1})}
\renewcommand{\sec}[1]{sec.\ \ref{#1}}

\newcommand{\fig}[1]{Fig.\ \ref{#1}}

\newcommand{\tr}{{\rm tr}}

\newcommand{\pic}[4]
{
 \begin{figure}
 \begin{center}
 \includegraphics[height=#3]{#4}
 \end{center}
 \caption{\label{#1} #2}
 \end{figure}
}

\newcommand{\qed}{\nobreak \ifvmode \relax \else
      \ifdim\lastskip<1.5em \hskip-\lastskip
      \hskip1.5em plus0em minus0.5em \fi \nobreak
      \vrule height0.75em width0.5em depth0.25em\fi}

\newcommand{\sixj}[6]{
\left\{
\begin{array}{ccc}
#1 & #2 & #3 \\ 
#4 & #5 & #6
\end{array}
\right\}
}

\newcommand{\ninej}[9]{
\left\{\begin{array}{ccc}
#1 & #2 & #3 \\
#4 & #5 & #6 \\
#7 & #8 & #9 
\end{array}\right\}
}

\newcommand{\bb}{\overline{b}}

\newcommand{\nablab}{\overline{\nabla}}

\newcommand{\mt}{\tilde{m}}

\begin{document}
\thispagestyle{empty}
\hfill
\parbox[t]{3.6cm}{
hep-th/yymmnnn \\
IGPG-06/10-5} 
\vspace{2cm}

\title{Analytic derivation of dual gluons and monopoles \\ from SU(2) lattice Yang-Mills theory \\ II.\ Spin foam representation}
\author{Florian Conrady}
\affiliation{Institute for Gravitational Physics and Geometry, Physics Department, Penn State University, University Park, PA 16802, U.S.A}
\email{conrady@gravity.psu.edu}
\preprint{IGPG-06/10-5}

\begin{abstract}
In this series of three papers, we generalize the derivation of dual photons and monopoles by Polyakov, and Banks, Myerson and Kogut,
to obtain approximative models of SU(2) lattice gauge theory. Our approach is based on stationary phase approximations.

In this second article, we start from the spin foam representation of 3-dimensional SU(2) lattice gauge theory. 
By extending an earlier work of Diakonov and Petrov, we approximate the expectation value of a Wilson loop by a path integral
over a dual gluon field and monopole-like degrees of freedom. The action contains the tree-level Coulomb interaction
and a nonlinear coupling between dual gluons, monopoles and current.
\end{abstract}

\maketitle

\section{Introduction}
\label{introduction}

It is well-known that in the strong-coupling regime of pure SU(N) gauge theory, confinement can be derived from an expansion in strong-coupling graphs \cite{Wilsonconfinement,KogutPearsonShigemitsu,Munsterhightemperature,DrouffeZuber}. It arises from three steps: an expansion in characters, an integration over the group variables, and an expansion in the inverse coupling. It is less known that the first two steps can be also done explicitly without any strong-coupling expansion: then, one obtains a non-perturbative sum over graphs that is exactly equivalent to the original lattice gauge theory \cite{Anishettyetal,HallidaySuranyi,OecklPfeifferdualofpurenonAbelian}. In this context, it would be misleading to speak of ``strong-coupling'' graphs, so we follow the paper \cite{OecklPfeifferdualofpurenonAbelian} and refer to them as \textit{spin foams}\footnote{This term was originally coined in the quantum gravity literature \cite{Baezspinfoammodels}.}. 

Physically, the spin foams may be viewed as worldsheets of flux lines. Thus, the spin foam representation is an approach to gauge theory where everything is formulated in terms of flux lines instead of fields. 

Up until recently, the strong-coupling expansion was the only example of an analytic computation with spin foams of gauge theory. Due to the strong coupling, fluctuations of the spin foam surfaces are strongly suppressed. The dominant contribution to the Wilson loop comes from the minimal surface that is spanned by the loop, and this gives the area law. The problem about this argument is that it requires a size of the lattice spacing that is comparable to the length scale of confinement. Therefore, it cannot serve as a completely satisfactory explanation. It would be preferable to have a derivation in the continuum limit that can produce both the short- and large-distance potential in a coherent fashion: i.e.\ with the Coulomb potential dominating at short distances, and the confining potential taking over at a certain length scale above the lattice cutoff.

This naturally suggests the following question: how can we perform an analytic computation with spin foams when the lattice spacing is small?
In the continuum limit, the spin foam surfaces undergo complicated fluctuations and it is not clear how one could sum over them. 

A main motivation for the string approach to gauge theory comes from a very similar question (for a review, see e.g. \cite{Polyakovgaugefieldsstrings,Polyakovliberation,Polyakovconfiningstrings,Antonovstringnature}): how can one find a continuum representation of gauge theory that implements the surface-like (or stringy) aspects of the strong-coupling expansion? 

The answer to our questions is known for the analogous problem of U(1) lattice gauge theory: there the counterpart of the spin foam formulation could be called a ``charge foam'' representation. It is given by a sum over branched surfaces that are labelled by U(1) representations (i.e. charges) and constrained by the Gauss law. In this context, the problem of confinement is solved by going to yet another representation: as was shown by Banks, Myerson and Kogut, the charge foam formulation can be transformed exactly to another representation that has dual photons and monopoles as its degrees of freedom. It was derived earlier by a different method by Polyakov \cite{PolyakovI,PolyakovII}. The photon-monopole representation allows for an analytic computation of the static potential between charges, and provides a direct explanation for confined and deconfined phases of the theory. For suitable values of the dimension and coupling, the monopoles condense along a string between the charges and thereby create a confining potential between them \cite{PolyakovII,BanksMyersonKogut,GopfertMack,Guth}.

The example of U(1) fosters the hope that one could generalize this scheme to non-abelian gauge theories: is there an analytic way to deduce analogues of the photon-monopole representation for SU(N) lattice gauge theory? In the three papers of this series, we address this question and derive approximative gluon-monopole models from three representations of SU(2) lattice Yang-Mills theory: from the BF Yang-Mills representation in dimension 3 and 4 \cite{ConradyglumonI}, from the spin foam representation in $d=3$, and from the plaquette representation in $d=3$ \cite{ConradyglumonIII}. 

In this paper, we will start from the spin foam representation in 3 dimensions and attempt to generalize the aforementioned transition from charge foams to 
dual photons + monopoles. We thereby extend an earlier work by Diakonov and Petrov \cite{DiakonovPetrov}, where it was already suggested how dual gluons arise from spin foams. 

The logic of the derivation is analogous to that of paper I: every step that was made within the BF Yang-Mills representation has a counterpart in the spin foam picture. It is non-trivial, however, to work out the details of this correspondence. The resulting model is similar to what we obtain in paper I and III: it contains a tree-level Coulomb interaction, and a nonlinear coupling between monopole-like excitations, dual gluons and current. 

The paper is organized as follows: after setting our conventions for SU(2) lattice gauge theory (\sec{SU2latticeYangMillstheoryin3dimensions}), we will review its spin foam representation in section \ref{spinfoamrepresentation}. In section \ref{relationbetweenBFYangMillstheoryspinfoamsumandgravity} we explain how notions of BF Yang-Mills theory translate into the spin foam representation. The main part is section \ref{gluonsandmonopolelikeexcitationsfromspinfoams}, where we derive the representation in terms of dual gluons and monopole-like excitations. The final section contains a summary and discussion of the results.

For conventions on lattice quantities, see the introduction of paper I \cite{ConradyglumonI}.

\section{SU(2) lattice Yang-Mills theory in 3 dimensions}
\label{SU2latticeYangMillstheoryin3dimensions}

The partition of function 3-dimensional SU(2) lattice Yang-Mills theory is defined by a path integral over SU(2)-valued link (or edge) variables $U_e$ on the lattice $\kappa$:
\be
\label{partitionfunction}
Z = \int\left({\ts\prod\limits_{e\subset\kappa^*}}\d U_e\right) \exp\Big(-\sum_f \clS_f(W_f)\Big)
\ee
The face (or plaquette) action $\clS_f$ depends on the holonomy $W_f$ around the face. As in paper I, we choose $S_f$ to be the heat kernel action (for more details on the definition, see \cite{MenottiOnofri}). The heat kernel action has a particularly simple expansion in terms of characters, namely, 
\be
\exp\Big(- \clS_f(W_f)\Big) = \sum_j\;(2j+1)\,\e^{-\frac{2}{\beta}\,j(j + 1)}\,\chi_j(W_f)\,.
\ee
The coupling factor $\beta$ is related to the gauge coupling $g$ via
\be
\beta = \frac{4}{a g^2} + \frac{1}{3}\,.
\ee

\section{Spin foam representation}
\label{spinfoamrepresentation}

\subsection*{Partition function}

\psfrag{Wilson loop}{Wilson loop $C$}
\pic{checkerboard}{Even cubes of the lattice $\kappa$.}{5cm}{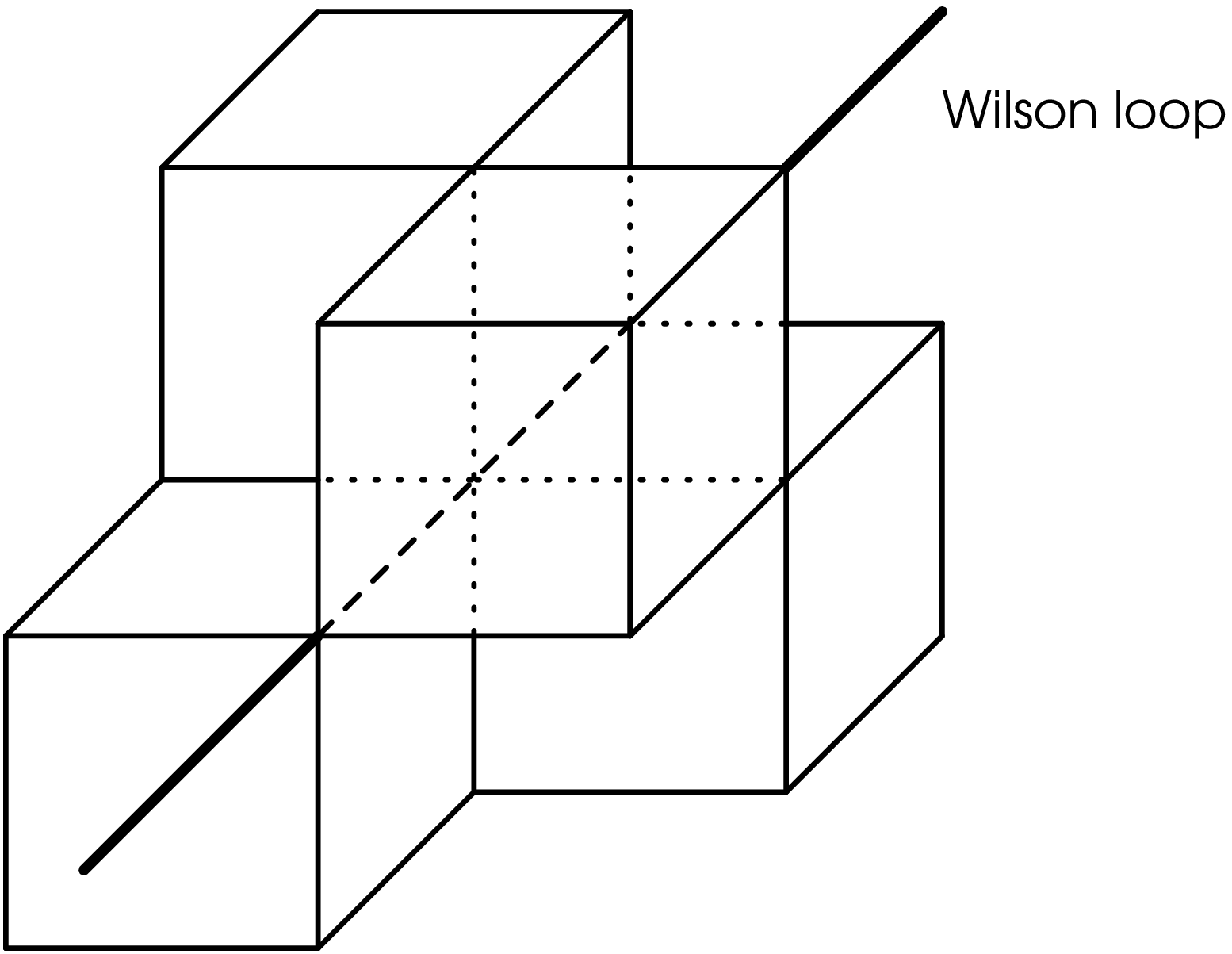}

The SU(2) lattice gauge theory can be rewritten in different representations. One of them is the representation in terms of spin foam sums (\cite{Anishettyetal,HallidaySuranyi,OecklPfeifferdualofpurenonAbelian}; for a review, see e.g.\ \cite{Conradygeometricspinfoams}). It results from the original path integral in two steps: first we expand the plaquette actions in gauge-invariant loops (a.k.a.\ characters), so that
\be
\label{firstorder}
Z = \int\left({\ts\prod\limits_{e\subset\kappa^*}}\d U_e\right) \sum_{\{j_f\}}\;
\left(\prod_{f\subset\kappa^*}\,(2j_f+1)\,\chi_{j_f}(W_f)\,\,\e^{-\frac{2}{\beta}\,j_f(j_f + 1)}\right)
\ee
We may view this as a first-order formulation, since it has two sets of variables: the connection variables $U_e$ and the spin assignments $j_f$. 

In the second step we integrate out the connection variables $U_e$. This can be done exactly and leads to a sum over configurations that we call spin foams. Each spin foam consists of an assignment of spins $j_f$ and intertwiners $I_e$ to faces and edges of $\kappa$ respectively. One may think of it as a branched surface that carries spin and intertwiner labels\footnote{Spin foams are essentially the same as the strong-coupling graphs of the strong-coupling 
expansion \cite{DrouffeZuber}. Their appearance, however, is not tied to any expansion in the coupling, so we prefer to use the term ``spin foam'', which was coined in the quantum gravity literature \cite{Baezspinfoammodels}. For a review of gravity spin foams, see e.g.\ \cite{Perezreview}.}. 

\psfrag{j1}{$\sst j_1$}
\psfrag{j2}{$\sst j_2$}
\psfrag{j3}{$\sst j_3$}
\psfrag{j4}{$\sst j_4$}
\psfrag{j5}{$\sst j_5$}
\psfrag{j6}{$\sst j_6$}
In general, there are several ways of organizing the integration steps, and each of them results in a different form of the spin foam sum.
Here, we will use an integration scheme that was proposed by Anishetty, Cheluvaraja, Sharatchandra and Mathur \cite{Anishettyetal}: it is specific to 3 dimensions, and leads to $6j$-symbols in the amplitude. In this case, the intertwiners are encoded by spin assignments $j_e$ to edges of $\kappa$, and the spin foams reduce to assignments of spins $j_f$ and $j_e$ to faces and edges respectively. To describe the amplitudes, it is convenient to use the following dual construction: imagine that we divide the cubes of $\kappa$ into sets of ``even'' and ``odd'' cubes, forming a ``checkerboard'' as indicated in fig.\ \fig{checkerboard}. Dually, we have a set of even and odd vertices in the dual lattice $\kappa^*$. Take the odd vertices, and connect each pair of odd vertices within the same face by a diagonal. In this way, the dual lattice $\kappa^*$ turns into a triangulation $T$ (see \fig{triangulation}). Faces of $\kappa$ are dual to edges of $\kappa^*$, and the edges of $\kappa$ are dual to the edges we added to $\kappa^*$ to get $T$. Thus, a spin foam can be described by assigning a spin $j_e$ to each edge $e$ of the triangulation $T$.  The spin foam sum takes the following form:
\be
\label{spinfoamsumpartitionfunction}
Z = \sum_{\{j_e\}_T}
\left(\prod_{e\subset T} (2j_e+1)\right) 
\left(\prod_{t\subset T} A_t\right)
\left(\prod_{e\subset\kappa^*}\,(-1)^{2j_e}\,\e^{-\frac{2}{\beta}\,j_e(j_e + 1)}\right)
\ee
Each configuration $\{j_e\}_T$ is an assignment of spins $j_e$ to edges $e$ of $T$ such that for each triangle of $T$ the spins satisfy the triangle inequality. 
The edges of $T$ belong to two groups: edges that are identified with edges in $\kappa^*$, and diagonal edges that were added to $\kappa^*$ in order to form the triangulation $T$. In the amplitude, every edge contributes with the dimension $2j_e + 1$ of the representaiton $j_e$. In addition, edges of $\kappa^*$ give a sign factor and an exponential of the Casimir. For each tetrahedron $t$, we get a $6j$-symbol 
\be
\label{amplitudetetrahedron}
A_t\quad =\quad \sixj{j_1}{j_2}{j_3}{j_4}{j_5}{j_6}
\quad =\quad 
\parbox{3.6cm}{\includegraphics[height=2.8cm]{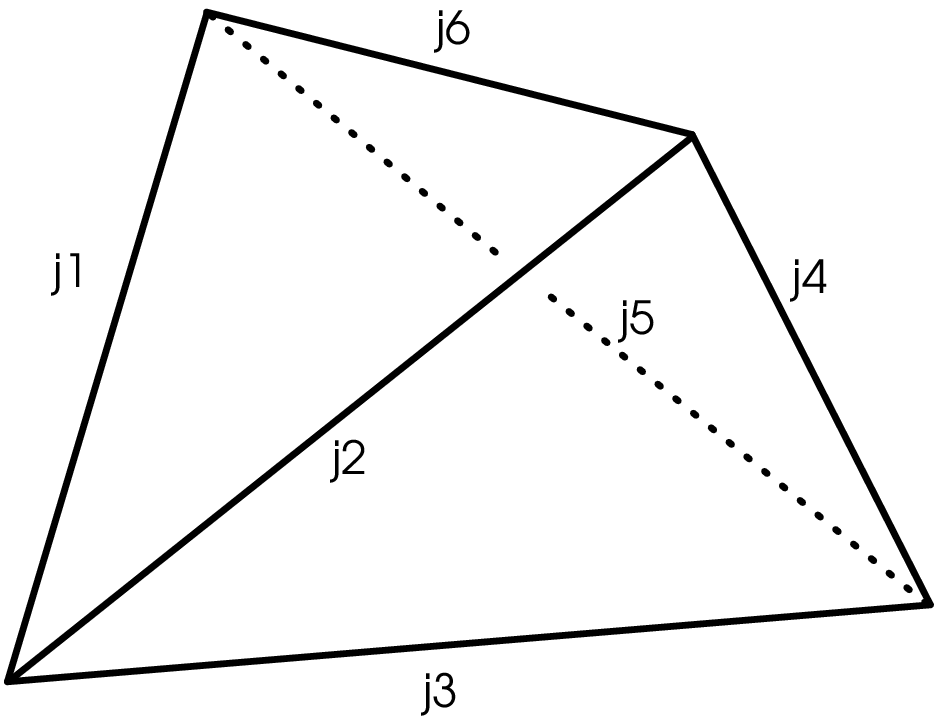}}\,.
\ee
where the spins $j_1$, $j_2$ and $j_3$ are read off from any triangle in the tetrahedron. The spins $j_4$, $j_5$, and $j_6$ are the spins on the edges opposing those of $j_1$, $j_2$ and $j_3$. Consider, for example, a tetrahedron that lies in the center of a dual cube: if we label its edges by spins as in fig.\ \fig{triangulation}, the associated amplitude is precisely expression \eq{amplitudetetrahedron}. 
The $6j$-symbol is equal to the value of a tetrahedral spin network (see the right-hand side of eq.\ \eq{amplitudetetrahedron}): the labelling of this spin network is dual to that of the tetrahedron in the sense that vertices of $t$ correspond to cycles in the spin network, and triangles of $t$ correspond to vertices\footnote{So that $6j$-symbol and spin network are equal, one has to specify suitable phase conventions for the intertwiners of the spin network.}.

\psfrag{j1}{$j_1$}
\psfrag{j2}{$j_2$}
\psfrag{j3}{$j_3$}
\psfrag{j4}{$j_4$}
\psfrag{j5}{$j_5$}
\psfrag{j6}{$j_6$}
\pic{triangulation}{Tetrahedra of the triangulation $T$ that form a cube of the dual lattice $\kappa^*$.}{6cm}{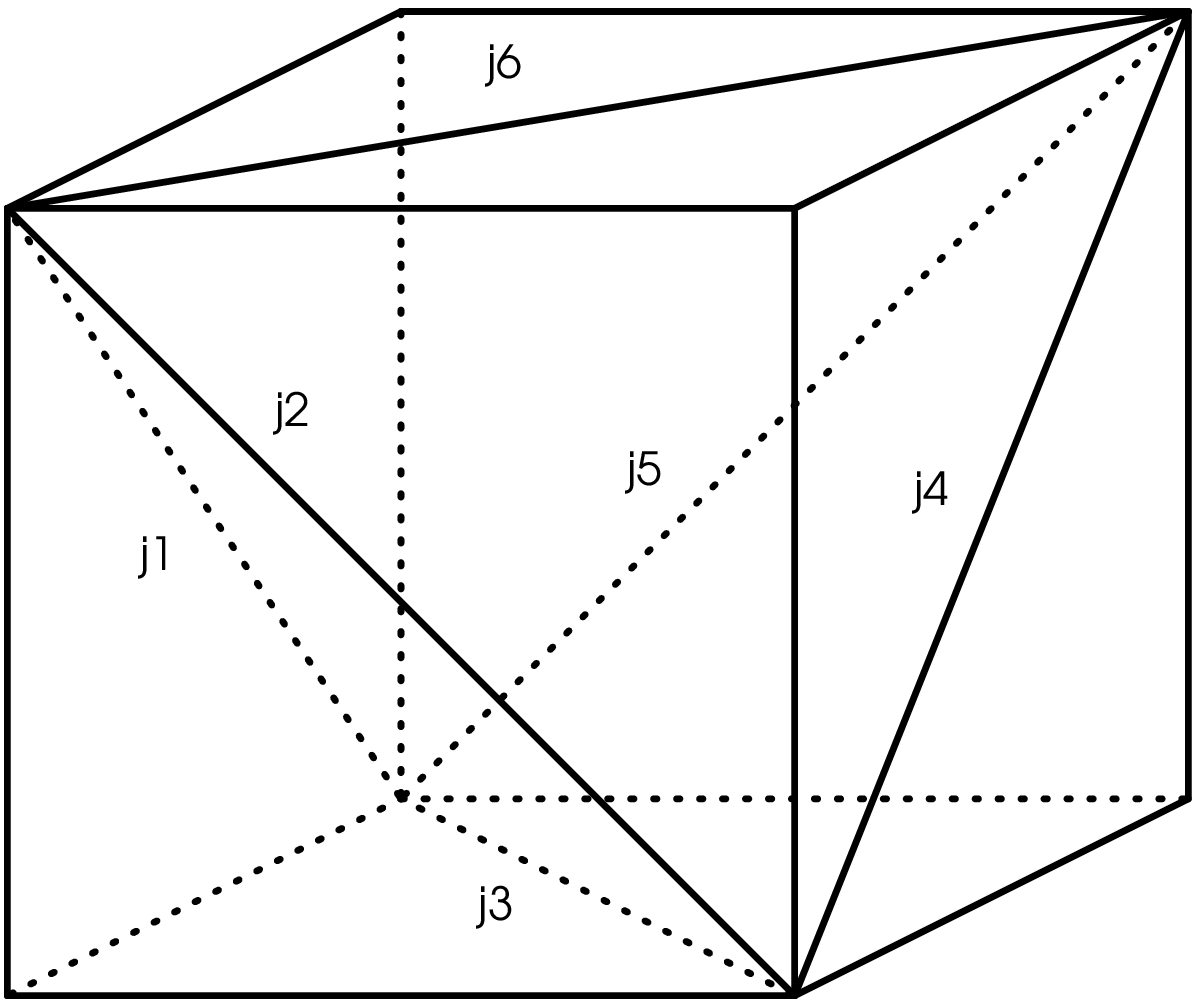}

\subsection*{Wilson loop}

How do we compute expectation values in the spin foam representation? Consider a Wilson loop $C$ in the representation $j$, where $C$ has no self-intersections. In the original formulation, the expectation value of $C$ is given by
\be
\label{Wilsonloop}
\b \tr_j W_C\ket = \int\left({\ts\prod\limits_{e\subset\kappa^*}}\d U_e\right) \,\exp\Big(-\sum_f \clS_f(W_f)\Big)\,\chi_j(W_C)\,.
\ee
$W_C$ denotes the holonomy around the Wilson loop. To describe the associated spin foam sum, we have to modify the triangulation $T$: consider all faces of the dual lattice $\kappa^*$ which are dual to edges of the Wilson loop. In each of these faces, we add a second diagonal edge. Together with the original diagonals, these edges define a ``hexahedron'' $h$ in the center of the cube. We call the resulting new complex $T'$ (see \fig{triangulationprime}). 

Then, the spin foam sum can be specified as follows:
\be
\label{expectationvalueofWilsonloop}
\b \tr_j W_C\ket = \frac{1}{Z}\,\sum_{\{j_e\}_{T'}}
\left(\prod_{e\subset T'} (2j_e+1)\right)
\left(\prod_{t\subset T'} A_t\right)
\left(\prod_{h\subset T'} A^{9j}_h\right) 
\left(\prod_{e\subset\kappa^*}\,(-1)^{2j_e}\,\e^{-\frac{2}{\beta}\,j_e(j_e + 1)}\right)
\ee
Let us explain the differences to formula \eq{spinfoamsumpartitionfunction}: each configuration $\{j_e\}_{T'}$ is an assignment of spins $e$ to edges of $T'$ such that 1.\ for each triangle of $T'$ the spins satisfy the triangle inequality, and 2.\ for every double diagonal the spins satisfy the inequality
\be
\left|j_i - j\right| \le j'_i \le j_i + j\,.
\label{triangleinequalitywithj}
\ee
In addition to $6j$-symbols, we now have $9j$-symbols---one for each hexahedron $h$ in $T'$ where the Wilson loop passes through\footnote{For details about sign factors, we refer the reader to Diakonov \& Petrov's paper \cite{DiakonovPetrov}.}:
\psfrag{j1}{$\sst j_1$}
\psfrag{j2}{$\sst j_2$}
\psfrag{j3}{$\sst j_3$}
\psfrag{j4}{$\sst j_4$}
\psfrag{j5}{$\sst j_5$}
\psfrag{j6}{$\sst j_6$}
\psfrag{j2p}{$\sst j'_2$}
\psfrag{j5p}{$\sst j'_5$}
\psfrag{j}{$\sst j$}
\be
\label{9jsymbol}
A^{9j}_h\quad =\quad \pm\ninej{j_1}{j_2}{j_3}{j'_5}{j}{j_5}{j_6}{j'_2}{j_4}
\quad =\quad
\pm\parbox{3.6cm}{\includegraphics[height=2.8cm]{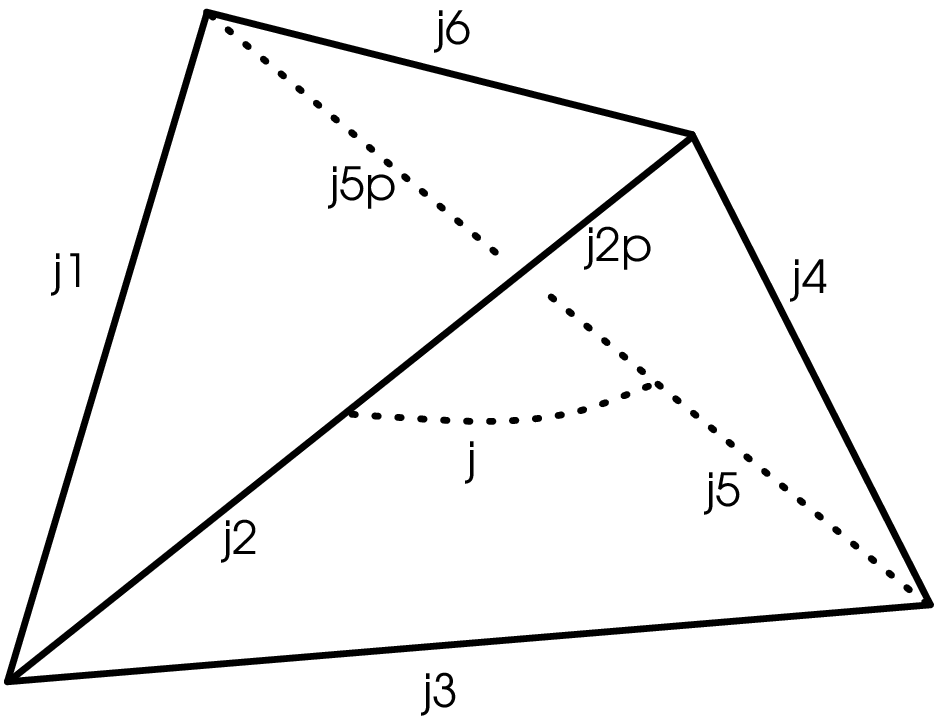}}
\ee 
The $9j$-symbol is equal to a spin network with an additional edge of spin $j$.

\psfrag{j1}{$j_1$}
\psfrag{j2}{$j_2$}
\psfrag{j3}{$j_3$}
\psfrag{j4}{$j_4$}
\psfrag{j5}{$j_5$}
\psfrag{j6}{$j_6$}
\psfrag{j2p}{$j'_2$}
\psfrag{j5p}{$j'_5$}
\pic{triangulationprime}{Modified triangulation $T'$ with double edges where the Wilson line passes through.}{6cm}{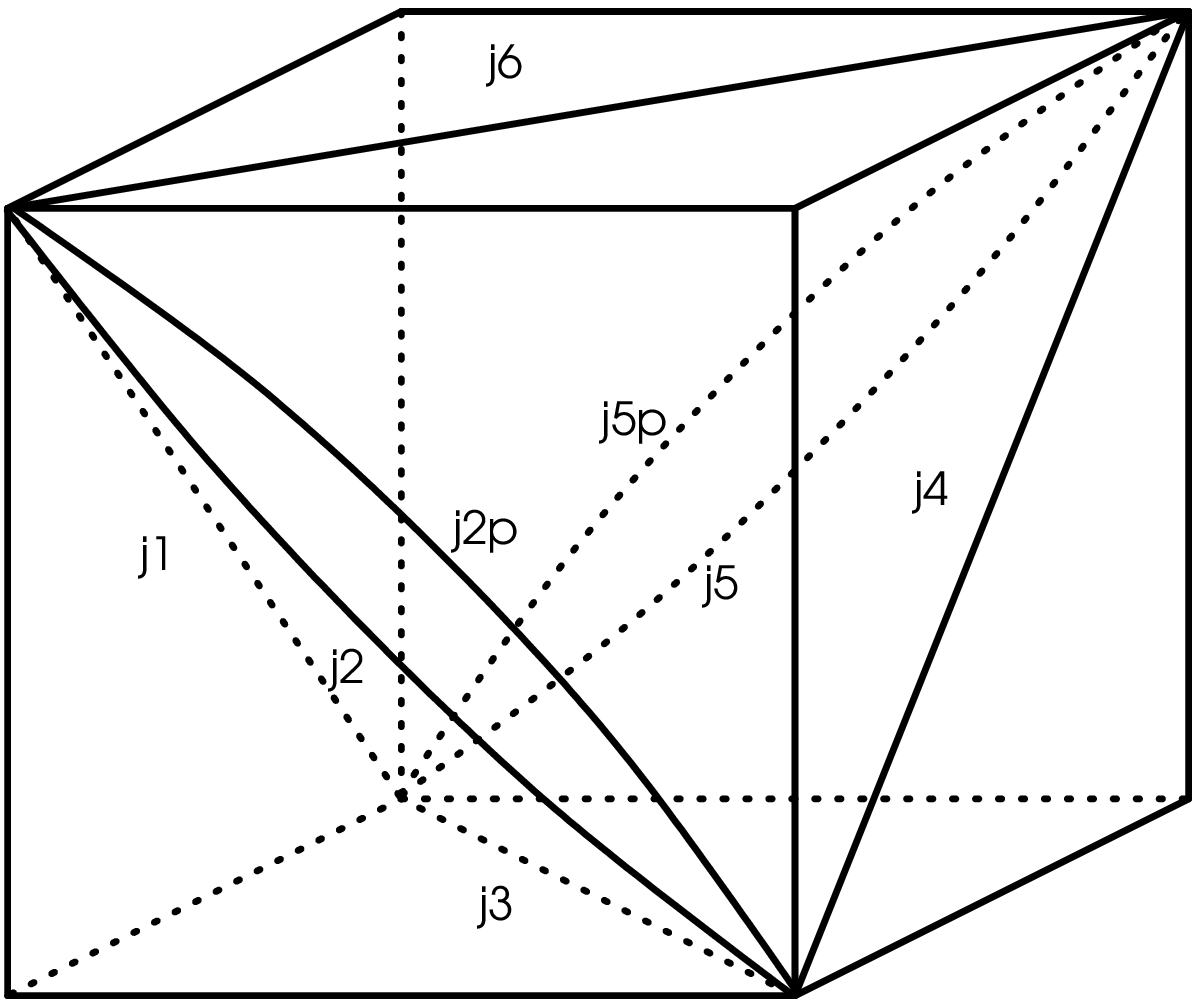}

\section{Relation between BF Yang-Mills theory, spin foam sum and gravity}
\label{relationbetweenBFYangMillstheoryspinfoamsumandgravity}

To derive an approximate model of the full lattice gauge theory, we use a similar argument as in paper I. This time, however, we will apply it at the level of the spin foam representation. In paper I, we considered the variation of the BF Yang-Mills action on the lattice, and argued that for high enough momenta $k$, stationary phase points are given by configurations where the connection is flat and the $B$-field satisfies a Gauss constraint. We proceeded \textit{as if} this was valid for all energy scales, and restricted the entire path integral to these stationary phase points: i.e.\ to flat connections and solutions of the Gauss constraint. The latter are expressed by $(d-3)$-forms on the dual lattice $\kappa^*$, so in $d=3$ we arrive at a path integral over a scalar field. 

To repeat this argument, we need to know how the variation of the $BF$-term and the $B^2$-term translates into the framework of the spin foam representation.
In paper I, we rewrote the SU(2) lattice Yang-Mills as a BF Yang-Mills theory on the lattice:
\bea
\lefteqn{Z_{\mathrm{YM}} = \int\left({\ts\prod\limits_{x\mu}}\;\d U_{x\mu}\right) 
\int_{\bR^3}\left({\ts\prod\limits_{x\mu\nu}}\;\d^3 b_{x\mu\nu}\,\sum_j\,\delta(|b_{x\mu\nu}| - j)\right)} \nonumber \\
&& \times\,\left({\ts\prod\limits_{x\mu\nu}}\,\frac{|\omega_{x\mu\nu}|/2}{\sin(|\omega_{x\mu\nu}|/2)}\right)
\exp\left[\sum_x\frac{\irm}{2}\,b_{x\mu\nu}\cdot\omega_{x\mu\nu} - \frac{1}{\beta}\, b^2_{x\mu\nu}\right]
\label{latticeBFYM}
\eea
It can be regarded as a deformation of BF theory on the lattice:
\bea
\lefteqn{Z_{\mathrm{BF}} = \int\left({\ts\prod\limits_{x\mu}}\;\d U_{x\mu}\right) 
\int_{\bR^3}\left({\ts\prod\limits_{x\mu\nu}}\;\d^3 b_{x\mu\nu}\,\sum_j\,\delta(|b_{x\mu\nu}| - j)\right)} \nonumber \\
&& \times\,\left({\ts\prod\limits_{x\mu\nu}}\,\frac{|\omega_{x\mu\nu}|/2}{\sin(|\omega_{x\mu\nu}|/2)}\right)
\exp\left[\sum_x\frac{\irm}{2}\,b_{x\mu\nu}\cdot\omega_{x\mu\nu}\right]
\eea
Both of these path integrals have an equivalent representation as a spin foam sum, namely
\bea
Z_{\mathrm{YM}} &=& \sum_{\{j_e\}}
\left(\prod_{e\subset T} (2j_e+1)\right) 
\left(\prod_{e\nsubseteq \kappa^*} (-1)^{2j_e}\right)
\left(\prod_{t\subset T} A_t\right)
\left(\prod_{e\subset\kappa^*}\e^{-\frac{2}{\beta}\,j_e(j_e + 1)}\right)\,,
\label{spinfoamsumBFYM} \\
Z_{\mathrm{BF}} &=& \sum_{\{j_e\}}
\left(\prod_{e\subset T} (2j_e+1)\right) 
\left(\prod_{e\nsubseteq \kappa^*} (-1)^{2j_e}\right)
\left(\prod_{t\subset T} A_t\right)\,.
\label{spinfoamsumBF}
\eea
We see that the difference lies in the deformation factor
\be
\label{dampening}
\exp\left(-\frac{2}{\beta}\,j_e(j_e + 1)\right)\,,
\ee
which is the spin foam analogue of the deformation term $B^2$ in \eq{latticeBFYM}. The BF-term, on the other hand, should be encoded in the tetrahedral amplitudes appearing in \eq{spinfoamsumBFYM} and \eq{spinfoamsumBF}. In order to make this correspondence concrete, we need to observe two things: 1.\ 3d SU(2) BF theory 
is equivalent to 3d first-order gravity. 2.\ In the large spin limit, the tetrahedral amplitudes are related to the action of 3d gravity.

In the continuum, the BF action is given by
\be
\label{naivecontinuumlimitBF}
\clS = \frac{1}{2}\int\d^3 x\;\epsilon_{\rho\mu\nu} B_{x\rho}\cdot F_{x\mu\nu}\,. 
\ee
The Lie algebras of SU(2) and SO(3) are isomorphic under the map (of basis elements)
\be
{\rm su(2)} \ni \irm\sigma^a/2\quad\mapsto\quad J^a \in {\rm so(3)}\,,\qquad a = 1,2,3,
\ee
where
\be
(J^a)^b{}_c := \epsilon^a{}^b{}_c\,.
\ee
Thus, we can interpret the forms $A^a$ and $F^a$ also as components of so(3)-valued forms $A$ and $F$. Then, we get
\be
F^b{}_c = F^a (J_a)^b{}_c = F^a\,\epsilon_a{}^b{}_c\,,
\ee
or
\be
F^a = \frac{1}{2}\,\epsilon^a{}_{bc}\,F^{bc}\,.
\ee
By inserting this in \eq{naivecontinuumlimitBF}, and identifying $B^a$ with the co-triad $e^a$, we see that the BF term is the same as the Hilbert-Palatini action of 3d gravity:
\be
\label{threedfirstordergravity}
\clS = \frac{1}{8}\int\d^3 x\;\;\irm\,\epsilon^{\rho\mu\nu} \epsilon_{abc}\, e^a{}_\rho F^{bc}{}_{\mu\nu}
\ee
Let us now consider the tetrahedral amplitudes in \eq{spinfoamsumBFYM}: they are given by $6j$-symbols
\be
A_t = \sixj{j_{e_1}}{j_{e_2}}{j_{e_3}}{j_{e_4}}{j_{e_5}}{j_{e_6}}\,.
\ee
Ponzano and Regge \cite{PonzanoRegge} interpreted the 6 spin assignments $j_e$ of a tetrahedron $t$ as lengths $j_e + 1/2$ of its edges $e$, and noticed that for non-degenerate tetrahedra and large spins the amplitudes
\be
A_t = \sixj{j_{e_1}}{j_{e_2}}{j_{e_3}}{j_{e_4}}{j_{e_5}}{j_{e_6}}
\ee
have the asymptotics
\be
\label{asymptotics}
A_t \approx \frac{1}{\sqrt{6\pi V}}\left(\e^{\irm R_t} + \e^{-\irm R_t}\right)\,.
\ee
Here, $V$ is the volume of the tetrahedron, and $R_t$ is the contribution of the tetrahedron to the Regge action---a discrete version of $\int\d^3 x \sqrt{g}\,R$ for piecewise flat geometries \cite{Regge}:
\be
\label{Reggeaction}
\sum_{t\subset T} R_t = \sum_{t\subset T} \left[\sum_{e\subset t} \left(j_e + 1/2\right)\theta_{te} + \pi/4\right]
\ee
$\theta_{te}$ stands for the dihedral angle at the edge $e\subset t$ in the tetrahedron $t$. 
This motivated Ponzano and Regge to propose the sum \eq{spinfoamsumBF} as a path integral quantization of 3-dimensional Euclidean gravity: 
the spin configurations describe 3d geometries and the $6j$-symbols provide the weighting with the gravity action.

In this way, we see how the $BF$-term in \eq{latticeBFYM} and the tetrahedral amplitudes in \eq{spinfoamsumBFYM} are related: 
both the BF-term and the tetrahedral amplitudes implement a weighting with a gravitational action. 
In the BF Yang-Mills path integral, we have the action of first-order Hilbert-Palatini gravity where the connection is an independent variable, while in the spin foam sum the connection is integrated out and the gravity term is that of second-order Einstein gravity.

%%%%%%%%%%%%%%%%%%%%%%%%%%%%%%%%%%%%%%%%%%%%%%%%%%%%%%%%%%%%%%%%%%%%%%%%%%%%%%%%%%%%%%%%%%%%%%%%%%%%%%%%%%%%%%%%%
%%%%%%%%%%%%%%%%%%%%%%%%%%%%%%%%%%%%%%%%%%%%%%%%% original part %%%%%%%%%%%%%%%%%%%%%%%%%%%%%%%%%%%%%%%%%%%%%%%%%
%%%%%%%%%%%%%%%%%%%%%%%%%%%%%%%%%%%%%%%%%%%%%%%%%%%%%%%%%%%%%%%%%%%%%%%%%%%%%%%%%%%%%%%%%%%%%%%%%%%%%%%%%%%%%%%%%

\section{Dual gluons and monopole-like excitations from spin foams}
\label{gluonsandmonopolelikeexcitationsfromspinfoams}

In this section, we start from the spin foam representation of the Wilson loop average, and derive an approximate representation
in terms of a dual gluon field and monopole-like variables. We thereby extend an earlier argument by Diakonov \& Petrov \cite{DiakonovPetrov}, 
where it was already suggested how dual gluons arise from spin foams. The line of reasoning stands in analogy to what we did in paper I within the 
 BF Yang-Mills representation. Technically, however, the present analysis is more complicated and we are forced to make various
 heuristic assumptions. The constraints, for example, give more solutions than in paper I---a feature that could be important for color screening. 
 
The derivation proceeds in four steps:
\begin{enumerate}
\item First we will apply the Poisson summation formula. We thereby trade the discreteness of spins for additional degrees of freedom which are
 similar to monopoles of U(1).
\item We use the asymptotic formula for $6j$ symbols.
\item By applying a stationary phase approximation we obtain a constraint on spin foams: 
it requires flatness {\it outside} and torsion {\it along} the Wilson loop. 
\item Solving the constraint yields the dual gluon degrees of freedom.
\end{enumerate}

\subsection*{Poisson summation formula}

Let us start from the spin foam sums of formula \eq{spinfoamsumpartitionfunction} and \eq{expectationvalueofWilsonloop}. We want to consider the continuum limit, so the lattice constraint $a$ is small and $\beta$ is much larger than 1. This means that the dampening due to the Casimir factor \eq{dampening} is weak, and that typical spins are much larger than 1/2. We therefore approximate
\be
2j_e+1 \approx 2j_e\,,\qquad j_e\;(j_e + 1) \approx j^2_e\,,\qquad \beta \approx \frac{1}{a g^2}\,.
\ee
A short calculation shows that we can apply the Poisson summation formula similarly as for U(1), giving us\footnote{We omit constants that will drop out in expectation values.}
\be
Z = 
\int_0^\infty\left({\ts\prod\limits_{e\in T}}\;j_e\,\d j_e\right)\,
\sum_{\{m_e\}}\,
\left(\prod_{t\subset T} A_t\right)
\left(\prod_{e\subset\kappa^*}\,(-1)^{2j_e}\,\e^{-\frac{2}{\beta}\,j^2_e}\right)\left(\prod_{e\subset T}\e^{4\pi\irm\,j_e m_e}\right)
\label{partitionfunctionafterPoisson}
\ee
and
\bea
\lefteqn{\b \tr_j W_C\ket = \frac{1}{Z}\,
\int_0^\infty\left({\ts\prod\limits_{e\in T'}}\;j_e\,\d j_e\right)\,
\sum_{\{m_e\}}} \nonumber \\
&& \times\,
\left(\prod_{t\subset T'} A_t\right)
\left(\prod_{h\subset T'} A^{9j}_h\right) 
\exp\left(-\irm \sum_{e\subset\kappa^*} 2\pi\,j_e  - \sum_{e\subset\kappa^*} \frac{2}{\beta}\,j^2_e + \sum_{e\subset T'} 4\pi\irm\,j_e m_e\right)\,.
\label{afterPoissonsummationformula}
\eea
It is assumed that we replace the discrete functions $A_t$ and $A^{9j}_h$ by suitable continuous functions of spin,
and that these functions have only support on spins that satisfy the triangle inequalities.

\subsection*{Stationary phase approximation}

Based on our previous considerations, we identify the $6j$ and $9j$ symbols in \eq{afterPoissonsummationformula} as the parts of the amplitude that correspond to the BF term and the source current of the Wilson loop. In order to determine the stationary phase points of the sum, we proceed in two steps: we will first consider the case without Wilson loop (i.e.\ the partition function \eq{partitionfunctionafterPoisson}) and then discuss how the stationary points are modified, when the Wilson loop is included. Without Wilson loop the argument is the same as in Diakonov \& Petrov's paper. We repeat it here for completeness.

If we assume that large spins dominate, the amplitudes of non-degenerate tetrahedra are described by the Ponzano-Regge formula \eq{asymptotics}. By plugging this into \eq{partitionfunctionafterPoisson}, we obtain
\bea
\lefteqn{Z = \int_0^\infty\left({\ts\prod\limits_{e\in T}}\;j_e\,\d j_e\right)\,\sum_{\{m_e\}}\,\sum_{\{s_t\}}} \nonumber \\
&& \hspace{-0.8cm}\times\,\left(\prod_{t\subset T} \frac{1}{\sqrt{6\pi V_t}} \exp\Big(\irm s_t R_t\Big)\right)
\exp\left(-\irm \sum_{e\subset\kappa^*} 2\pi\,j_e 
- \sum_{e\subset\kappa^*} \frac{2}{\beta}\,j^2_e + \sum_{e\subset T} 4\pi\irm\,j_e m_e\right)\,.
\label{partitionfunctionafterasymptoticformula}
\eea
We sum over signs $s_t = \pm 1$ for each tetrahedron, due to the two terms of opposite phase in formula \eq{asymptotics}.

\setlength{\jot}{0.3cm}
Let us try to determine the stationary phase points of this expression. If we ignore the polynomial volume factors, the relevant exponent is
\bea
\clS &=& \irm \sum_{t\subset T} s_t \left[\sum_{e\subset t} j_e\,\theta_{te} + \pi / 4\right] 
- \irm \sum_{e\subset\kappa^*} 2\pi\,j_e - \sum_{e\subset\kappa^*} \frac{2}{\beta}\,j^2_e + \sum_{e\subset T} 4\pi\irm\,j_e m_e \nonumber \\
&=& \irm \sum_{e\subset T} j_e\,\Theta_e 
+ \sum_{e\subset T} \pi / 4 - \sum_{e\subset\kappa^*} \frac{2}{\beta}\,j^2_e + \sum_{e\subset T} 4\pi\irm\,j_e m_e\,,
\label{exponent}
\eea
where we set $j_e + 1/2 \approx j_e$ and defined the angle
\be
\Theta_e = \sum_{t\supset e} s_t\,\theta_{te} - \bigg\{
\parbox{4cm}{\renewcommand{\arraystretch}{1.3}
$\begin{array}{ll}
2\pi\,, & e\subset\kappa^*\,, \\
4\pi\,, & e\nsubseteq\kappa^*\,.
\end{array}$}
\ee
The sum over alternating signs $s_t$ leads to rapid phase oscillations, so we expect that a stationary phase point can be only reached if the signs $s_t$ are either all $+1$ or all $-1$. In that case, $\Theta_e$ is the deficit angle at the edge $e$: an edge in $\kappa^*$ is contained in 4 tetrahedra, and a diagonal edge lies in 6 tetrahedra; hence, for a flat geometry, the sum of dihedral angles at $e$ is $4\pi - 2\pi = 2\pi$ and $6\pi - 2\pi = 4\pi$ respectively.
\setlength{\jot}{0cm}

The first term in \eq{exponent} corresponds to the continuum expression $\sqrt{\det g}\,R$. In the spin foam geometry, lengths are given by spins, so the metric is of the order $g \sim j^2$. Since the Ricci scalar $R$ goes roughly like $g^{-1} g^{-1} \pa\pa g$, this gives us
\be
j\,\Theta \sim \sqrt{\det g}\,R \sim \sqrt{\det g}\,g^{-1} g^{-1} \pa\pa g \sim (j^6)^{\frac{1}{2}} j^{-2} j^{-2} \pa\pa j^2 \sim \pa\pa j\,.
\ee
For Fourier modes $j(k)$ of the spin field, where $k$ is sufficiently high, the phase oscillations in \eq{exponent} should be dominated by 
the curvature term. Therefore, a stationary phase point is reached when $\Theta_e$ is zero for all edges, i.e.\ when the spin foam geometry is flat. 
This is analogous to the argument in paper I, where we concluded that for high momenta stationarity is attained by flat connections. Like there, we will proceed \textit{as if} this was valid for the whole momentum range, i.e.\ we will restrict the entire sum \eq{partitionfunctionafterPoisson} to flat spin foams.

How does the situation change, when we include the Wilson loop? Then, the analysis is less clear, since we have no asymptotic formula for the $9j$-symbols associated to ``hexahedra''. We know, however, that the spins $j_i$ are typically larger than the fixed $j$ of the source, and that the $9j$-symbol is zero unless $j'_i$, $j_i$ and $j$ satisfy the triangle inequality (see eq.\ \eq{9jsymbol}). The simplest scenario would be that the $9j$-symbol has just the effect of ensuring the triangle inequality, and can be otherwise approximated by $6j$-symbols: that is, if
\be
\label{additionalcondition}
|j_i - j|\; \le\; j'_i\; \le\; j_i+j\,,
\ee
the modulus
\be
\left|A^{9j}_h\right| \quad=\quad \left|\ninej{j_1}{j_2}{j_3}{j'_5}{j}{j_5}{j_6}{j'_2}{j_4}\right|
\ee
is approximated by the {\itshape asymptotics} of 
\be
\left|\sixj{j_1}{j_2}{j_3}{j_4}{j_5}{j_6}\right|\,,
\ee
and otherwise
\be
A^{9j}_h\quad =\quad 0\,.
\ee
If this is true, what will be the stationary points? Far from the Wilson loop the stationary spin foams should be flat according to our previous argument.
Directly at the Wilson loop this argument does not apply, since we do not know the influence of the phase of $A^{9j}_h$.
We only know that the additional condition \eq{additionalcondition} has to be satisfied.

Since we lack a more precise analysis, we will assume that the stationary points are given by {\it all} spin foams that are flat outside of $C$ and 
meet condition \eq{additionalcondition} along the Wilson loop. By ``flat outside of $C$'' we mean the following property: one can
\begin{itemize}
\item remove the union $H$ of all ``hexahedra'' $h$ along $C$ from $T'$, and
\item map the remaining spin foam into Euclidean $\bR^3$ such that
\item the induced metric on $T' \backslash H$ is flat, and
\item spin assignments coincide with edge lengths determined by this metric.
\end{itemize}
Such a map sends every edge $e$ of $T'$ into a difference vector $b_e$ in $\bR^3$ and the length of $b_e$ agrees with the spin:
\be
|b_e| = \sqrt{b^a_e b^a_e} = j_e + 1/2
\ee
Note that for any closed curve around a triangle, the corresponding image vectors $b_e$ close as well. If we view the $b_e$'s as a 2-chain $b$ on $T'$, it satisfies $\d b = 0$ for all faces outside the dual $*C$. Around faces $f\subset T'$ that are dual to $C$ the vectors $b_e$ do \textit{not} close, in general, since $b_e$ and $b_{e'}$ may have different lengths (see \fig{deficit}). They are only restricted by the triangle inequality \eq{additionalcondition}:
\be
\label{triangleinequalityforb}
\Big||b_e| - j\Big| \quad \lesssim \quad |b'_e|\quad \le\quad |b_e| + j
\ee
We can regard this failure of vectors to close as a defect in our otherwise flat spin foam geometry:
the technical term for this deviation is \textit{torsion}, and the 2-form $\d b$ measures the amount of torsion in the spin foam.
The Wilson loop $C$ represents the worldlines of quarks, and that we get nonzero torsion along them fits well
with the fact that classically a fermion current is a source of torsion \cite{HehlHeydeKerlick}.

\psfrag{T'}{$\sst T'$}
\psfrag{R3}{$\sst \bR^3$}
\psfrag{f}{$\sst f$}
\psfrag{e}{$\sst e$}
\psfrag{be}{$\sst b_e$}
\psfrag{e'}{$\sst e'$}
\psfrag{be'}{$\sst b'_e$}
\psfrag{Jf}{$\sst J_f$}
\pic{deficit}{Torsion vector $J_f$ on faces $f$ dual to $C$.}{4cm}{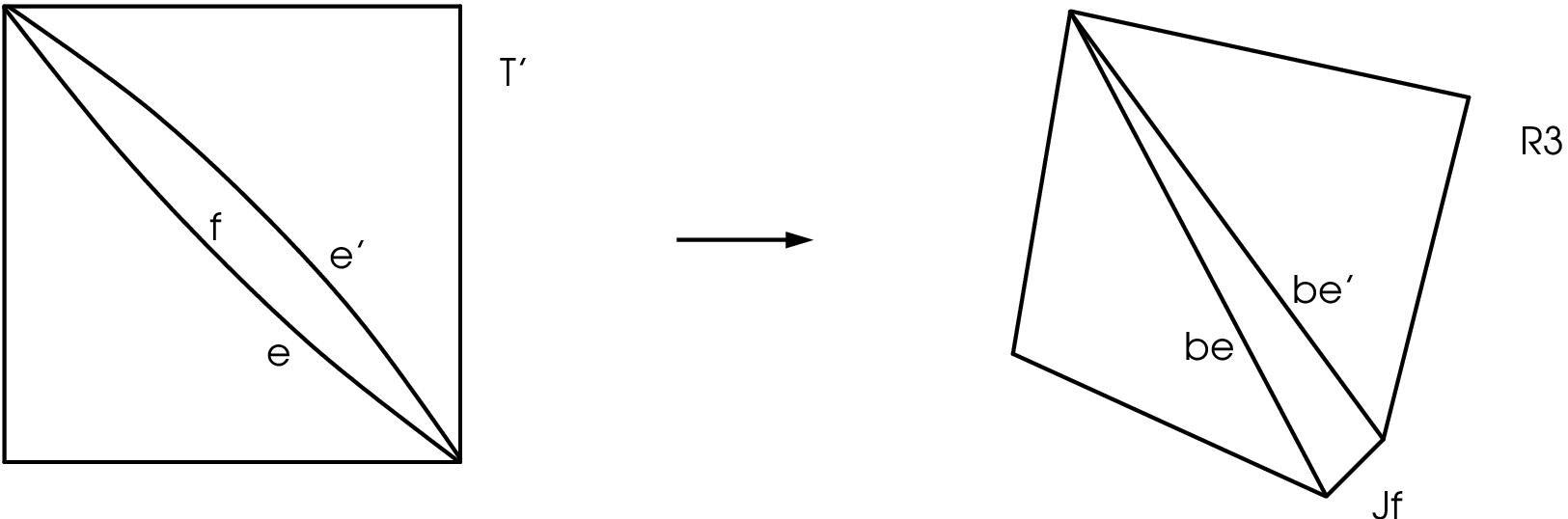}

The torsion along the loop is constrained by the triangle inequality. We can satisfy it by interpreting the spin $j$ in 
\eq{triangleinequalityforb} as the length of a third vector, and make all three vectors sum up to zero. More precisely, condition \eq{triangleinequalityforb} 
is met if there is a vector
\be
J = j\,n\,,\qquad n\in\bR^3,\;\; |n| = 1\,,
\ee
such that either 
\be
\mathrm{(I)}\;\; b_e - b_{e'} = J\,,\qquad\mbox{or}\qquad \mathrm{(II)}\;\; b_e + b_{e'} = J\,.
\ee 
Suppose that the edges $e$ and $e'$ are oriented such that $\pa f = e - e'$. Then, possibility (I) means that
\be
\label{abelianGausslaw}
\d b_f = J\,,
\ee
and the magnitude of torsion is determined by the spin $j$ of the Wilson loop. The equation looks much like the abelian
Gauss constraint of U(1) lattice gauge theory. Solution (II), on the other hand, leads to a torsion 
\be
\label{violationofGausslaw}
\d b_f = b_e - b_{e'} = 2 b_e - J \approx 2 b_e\,,
\ee
and the abelian Gauss law is violated. In that case, the torsion would be of the order of the typical spins in the spin foams, which is large.
In the following, we will drop possibility (II) and only work with the solution (I). 
There are two motivations for this: the first one is simply that we do not know how to analyze condition (II).
Secondly, we are not sure that the solution (II) would appear in an exact analysis of $9j$-symbols.
We would expect that the magnitude of torsion is determined by the representation of the fermions, and not by the representation of typical fluctuations.

Even with this assumption the Gauss law \eq{abelianGausslaw} is not the only way of solving the triangle inequality.
The triangle inequality states a condition on the length of the vectors $b_e$ and $b_{e'}$, but does not necessarily imply that there is
a third vector of length $j$ with which they sum up to zero. We can, for example, take $b_e$ and $b_{e'}$ to be the same and meet the inequality. 
If we combine this with the condition $*\d b = 0 = \pa\! *\!b$ outside the loop, we can construct a solution by 
using a tube-like surface $S$ that ends on the Wilson loop $C$, and carries some constant vector $v\in\bR^3$ as ``color charge'':
\be
\label{colorscreeningsolution}
*b = v\,S
\ee
It corresponds to a spin foam with surface $S$ and uniform spin $|v|$. 
Recall now that the sum over monopole-like variables has the effect of restraining spins to be integers or half-integers. 
As a result, the spin foam surface \eq{colorscreeningsolution} is suppressed unless $|v| \in \bZ_+/2$. Then, however, the triangle inequality
\be
||v| - j| \quad \le \quad |v|\quad \le\quad |v| + j
\ee
is only met if the spin $j$ is \textit{integer}! In other words, the tube-like configuration \eq{colorscreeningsolution} is only allowed when the Wilson loop has integer spin.

Such type of spin foams are precisely the diagrams that are responsible for color screening in the strong-coupling regime. We suspect them to be also the cause for color screening in the continuum limit. Our argument shows that these spin foams appear as solutions to our flatness and defect condition. We also see that the Gauss law \eq{abelianGausslaw} has a smaller set of solutions and cannot capture this non-abelian effect.

For simplicity, we will not deal with these additional solutions and instead just work with the naive condition \eq{abelianGausslaw}.
In that case, the deviation from $\d b = 0$ can be described by an $\bR^3$-valued 2-chain $J$ that has length $j$ along $*C$, and is otherwise zero.
Thus, we can express the flatness and torsion condition by a single condition on the 2-chain $b$:
\be
\label{constraint}
\d b = J\,.
\ee
Since $C$ has no self-intersections and $\d J = \d^2 b = 0$, the vector $J_f$ is always the same along $C$, i.e.\
\be
\label{source}
J_f \quad = \quad 
\Bigg\{\renewcommand{\arraystretch}{1.8}
\parbox{3.5cm}{
$\begin{array}{ll}
j\,n\,, & \mbox{$f$ dual to $C$}\,,\\
0\,, & \mbox{otherwise}\,,
\end{array}$}
\ee
for some unit norm vector $n$. This provides the desired characterization of stationary points: a spin foam is stationary if there is an $\bR^3$-valued 1-chain $b$ on $T'$ and a current $J$ of the form \eq{source} such that $\d b = J$ and $j_e = |b_e|$ for all edges $e\subset T'$. 

We now constrain the path integral \eq{afterPoissonsummationformula} to these stationary phase points\footnote{We will ignore additional factors that would come from integrating over the quadratic order in fluctuation variables.}. 
The constraint has the general solution 
\be
b = \d\varphi + \bb\,,
\ee
where $\varphi$ is an $\bR^3$-valued scalar field on the dual lattice $\kappa^*$ and $\bb_{x\mu}$ is a particular solution.
The latter can be chosen as
\be
\bb = j\,n*\!S
\label{surface}
\ee
for any surface $S$ bounded by $C$. 
Thus, we can make a change of variables
\be
\int_0^\infty\left({\ts\prod\limits_e}\;j_e\,\d j_e\right) \quad \ldots
\quad\to\quad  
\int_{\bR^3}\left({\ts\prod\limits_v}\;\d^3\varphi_v\right) |\clJ| \quad \ldots\,,
\ee
and express the constrained path integral by an unconstrained path integral over the scalar field.
The Jacobian is given by
\be
\clJ = \det\left(\frac{1}{2}\pdiff{j^2_e}{\varphi^a_v}\right)\,,
\quad \mbox{where}\qquad 
\frac{1}{2}\pdiff{j^2_{e}}{\varphi^a_v}\; =\; \left(\d\varphi^a_v + \bb^a\right) \times 
\left\{\renewcommand{\arraystretch}{1.6}
\raisebox{0cm}[0.8cm][0.8cm]{
$\begin{array}{rl}
1\,, & \mbox{$e$ incoming at $v$}\,,\\
-1\,, & \mbox{$e$ outgoing at $v$}\,, \\
0\,, & \mbox{otherwise}\,.
\end{array}$}\right.
\ee
When making this change of variables, we have to observe that the map from $\varphi$ to $j$ has global degeneracies: 
a global translation and rotation of the values of $\varphi$ in $\bR^3$ does not affect the spins.
We fix the translational degeneracy by removing the Fourier mode of momentum zero from the $\varphi$-integration.
The rotational invariance produces a finite and field-independent volume factor (the volume of the rotation group) and drops out 
in expectation values. With this, the path integral over $\varphi$ takes the form
\bea
\lefteqn{\b \tr_j W_C\ket = \frac{1}{Z}\,
\int_{\bR^3}\left({\ts\prod\limits_v}\;\d^3\varphi_v\right)' |\clJ|\,
\sum_{\{m_e\}}\,\frac{1}{4\pi}\int_{S^2}\d n} \nonumber \\
&& \times\,
\left(\prod_{t\subset T'} A^{6j}_t\right)
\left(\prod_{h\subset T'} A^{9j}_h\right) 
\exp\left(-\irm \sum_{e\subset\kappa^*} 2\pi\,j_e - \sum_{e\subset\kappa^*} \frac{2}{\beta}\,(\d\varphi_e + \bb_e)^2 + \sum_{e\subset T'} 4\pi\irm \left|\d\varphi_e + \bb_e\right| m_e\right)\,. \nonumber \\
\eea
We have included an integration over the vector $n$ in $\bb$, since its choice is arbitrary.

The restriction to flat spin foams simplifies the contribution from $6j$- and $9j$-symbols. Recall that the asymptotic expression for $6j$-symbols contains volume factors $V_t$ and phase factors $\exp(\irm R_t)$. The latter combined with the sign factors $(-1)^{2j_e}$ to give deficit angles $\Theta_t$ in the exponent. In our integral over flat spin foams, these deficit angles are all zero. For simplicity, we will also drop all factors with a polynomial dependence on spins, i.e.\ we remove both the Jacobian $\clJ$ and the volume factors $V_t$. This means that the contribution from $6j$-symbols is effectively 1. We approximated the $9j$-symbols by $6j$-symbols, provided the triangle inequality \eq{additionalcondition} is satisfied. Thus, the $9j$-amplitudes become trivial as well. 

Before we write down this approximation, we change to a notation in terms of indices---writing $(x\mu)$ instead of $e$ and $(x\mu\nu)$ in place of $f$.
The $m_e$'s that were associated to edges of $\kappa^*$ are designated by $(x\mu)$, i.e.\ we write $m_{x\mu}$.
The $m_e$'s that were associated to diagonal edges are designated by the plaquette $(x\mu\nu)$ in which the diagonal edge lies, and we write them as $\mt_{x\mu\nu}$. The final result is
\bea
\lefteqn{\b \tr_j W_C\ket = 
\frac{1}{Z}\int_{\bR^3}\left({\ts\prod\limits_x}\;\d^3\varphi_x\right)'\sum_{\{m_{x\rho}\}}\sum_{\{\mt_{x\rho\sigma}\}}\;\frac{1}{4\pi}\int_{S^2}\d n} \nonumber \\
&& \times\,
\exp\left[\sum_x
\left(
-\frac{2}{\beta}\left(\nablab_\mu\varphi_x + \bb_{x\mu}\right)^2 \right.\right.\nonumber \\
&& \hspace{2.2cm}{}+ 4\pi\irm\,\left|\nablab_\rho\varphi_x + \bb_{x\rho}\right| m_{x\rho} \nonumber \\
&& \hspace{2.2cm}{}+ 4\pi\irm\,
\left|\nablab_\rho\varphi_x + \bb_{x\rho}
+ \nablab_\sigma\varphi_x + \bb_{x\sigma}
\right|\mt_{x\rho\sigma}
\bigg)\Bigg]\,.
\label{finalresult}
\eea 
In the last term of the exponent, the repeated indices $\rho$ and $\sigma$ are only summed over the pairs $\rho < \sigma$.
This expression is very similar to what we found in paper I. There are only three differences:
\begin{itemize}
\item Here, we have a second set of monopole variables $\mt$ which arise from the discreteness of spins on diagonal edges.
\item The norm of $J$ is $j$, while it is $j+1/2$ in paper I.
\item The overall normalization is not clearly determined by the present derivation. 
\end{itemize}
As in paper I, we can factor off the Coulomb interaction between the currents: for this purpose, we choose the particular solution as
\be
\bb_{x\rho} = -\epsilon_{\rho\mu\nu} u_\mu\,(u\cdot\nabla)^{-1} J_{x\nu}\,.
\ee
Then, after changing variables
\be
\varphi_x - \Delta^{-1}\nabla_\mu\bb_{x\mu}\quad\rightarrow\quad\varphi_x\,,
\ee
and using the identity
\be
\label{identityforJ}
\nabla_\mu\bb^a_{x\mu}\Delta^{-1}\nabla_\mu\bb^a_{x\mu} + \bb^2_{x\mu} = - J^a_{x\mu}\Delta^{-1}J^a_{x\mu}\,,
\ee
we obtain \setlength{\jot}{0.3cm}
\bea
\lefteqn{\b \tr_j W_C\ket = 
\frac{1}{Z}\int_{\bR^3}\left({\ts\prod\limits_x}\;\d^3\varphi_x\right)'\sum_{\{m_{x\mu}\}}\sum_{\{\mt_{x\mu\nu}\}}\;\frac{1}{4\pi}\int_{S^2}\d n} \nonumber \\
&& \times\,
\exp\Bigg[\sum_x\bigg(\frac{2}{\beta}\,\varphi^a_x\Delta\varphi^a_x 
+ 4\pi\irm\left|\nablab_\mu\left(\varphi_x + \Delta^{-1}\nabla_\nu\bb_{x\nu}\right) + \bb_{x\mu}\right|m_{x\mu} \nonumber \\
&& \hspace{2.2cm}{}+ 4\pi\irm\,\left|\nablab_\mu\left(\varphi_x + \Delta^{-1}\nabla_\rho\bb_{x\rho}\right) + \bb_{x\mu} \right. \nonumber \\
&& \hspace{3.4cm}{}+ \left.\nablab_\nu\left(\varphi_x + \Delta^{-1}\nabla_\sigma\bb_{x\sigma}\right) + \bb_{x\nu}\right|\mt_{x\mu\nu} \nonumber \\
&& \hspace{2.2cm}{}+ \frac{2}{\beta}\,J^a_{x\mu}\Delta^{-1}J^a_{x\mu}
\bigg)\Bigg]\,.
\label{result}
\eea
\setlength{\jot}{0cm}

\section{Summary and discussion}
\label{summaryanddiscsussion}

In this paper, we have derived an approximative model for 3-dimensional SU(2) lattice gauge theory. Its degrees of freedom can be interpreted as 
a dual gluon and monopole-like field. We propose it as a generalization of the photon-monopole representation of Polyakov \cite{PolyakovI,PolyakovII} and Banks, Myerson and Kogut \cite{BanksMyersonKogut}.

The derivation starts from the exact spin foam representation of SU(2) lattice gauge theory and extends an earlier work of Diakonov and Petrov \cite{DiakonovPetrov}. The line of argument and the result are similar as in paper I. The crucial step is a stationary phase approximation: it can be viewed as the zeroth order of a weak-coupling perturbation theory that has two types of non-trivial field configurations as a background: 1.\ the monopole-like excitations that arise from the compactness of the gauge group, and 2.\ the particular solution of the Gauss constraint which carries information about the large-distance defect created by the current.

The resulting model contains two interaction terms: firstly, a current-current potential that is essentially the tree-level Coulomb interaction one would get from a purely perturbative treatment. Secondly, a coupling between monopole-like excitations, dual gluons and current. This coupling is similar to the photon-monopole coupling of U(1), but nonlinear.  

At the technical level, the present derivation is more complicated than in the first paper, and we have to make several heuristic arguments.
To make this more precise, we would need the asymptotics of $9j$-symbols, or another, systematic way of analyzing the sum over spin foam amplitudes.

In spite of its heuristic character, the derivation from the spin foam representation is particularly interesting: 
it is the representation in which the strong-coupling expansion is performed, and in this regime confinement is a direct consequence
of the spin foam diagrams. By analyzing the same representation at weak coupling we can ask the question how this stringy behaviour is preserved or modified as we 
go to the continuum limit. 

As we translated the steps of paper I to the spin foam representation, we found more solutions to the stationary phase condition than in the first paper. 
We learned that the Gauss constraint of the form \eq{abelianGausslaw} covers only a subset of solutions and cannot capture all aspects of the non-abelian theory. When the spin of the Wilson loop is integer, the larger set of solutions includes  spin foams of the type that are responsible for color screening at strong coupling. We suspect them to be also the cause for color screening in the continuum limit.

The model of this paper can be tested by simulations and compared to simulations of the full theory. By summing over the monopole variables in expression \eq{finalresult} one can remove the phase factors and translate them into a constraint. The latter can be enforced by a Gaussian damping factor.

\section*{Acknowledgements}

I thank Abhay Ashtekar, Gerhard Mack, Alejandro Perez and Hendryk Pfeiffer for discussions. 
This work was supported in part by the NSF grant PHY-0456913 and the Eberly research funds. 

\bibliography{bibliography}
\bibliographystyle{hunsrt}  

\end{document}